\documentclass[prmaterials,superscriptaddress,aps,preprint,showkeys,longbibliography,floatfix]{revtex4-2}

\usepackage{amsmath,amssymb,txfonts}
\usepackage[version=3]{mhchem} 
\usepackage{multirow}
\usepackage{graphicx}
\usepackage{makecell}
\usepackage{hyperref}
\usepackage{xcolor}

\date{\today}

\begin{document}

\title{Investigating the effects of local environment on nitrogen vacancies in high entropy metal nitrides}
\author{Charith R. DeSilva}
    \affiliation{Materials Science \& Engineering Department, University of Illinois at Urbana-Champaign}
\author{Matthew D. Witman}
    \affiliation{Sandia National Laboratories, Livermore, CA}
\author{Dallas R. Trinkle}
    \email{dtrinkle@illinois.edu}
    \affiliation{Materials Science \& Engineering Department, University of Illinois at Urbana-Champaign}

\begin{abstract}

High entropy metal nitrides are an important material class in a variety of applications, and the role of nitrogen vacancies is of great importance for understanding their stability and mechanical properties.
We study six different high entropy nitrides with eight different metal species to build a predictive model of the nitrogen vacancy formation energy.
We construct sets of supercells that maximize the number of unique nitrogen environments for a given chemistry, and then use density-functional theory to calculate the energy density for all nitrogen sites, and the vacancy formation energies for the highest, lowest, and a median subset based on the energy densities.
The energy density of nitrogen sites correlates with the vacancy formation energies, for binary, ternary and high entropy nitrides.
A linear regression model predicts the vacancy formation energies using only the nearest-neighbor composition; across our eight metals, we find the largest vacancy formation energies next to Hf, then Zr, Ti, V, Cr, Ta, Nb, and the lowest near Mo. 
Additionally, we see that binary nitride data shows qualitatively similar vacancy formation energy trends for high entropy nitrides; however, the binary data alone is insufficient to predict the complex nitride behavior.
Our model is both predictive and easily interpretable, and correlates with experimental data. 

\end{abstract}

\maketitle

\section{Introduction}

Since the concept of high entropy alloys was first introduced in 2004 \cite{cantor2004microstructural,yeh2004nanostructured}, research interest in high entropy materials has expanded rapidly to include high entropy ceramics \cite{oses2020high}.
High entropy ceramics are defined as single-phase interstitial solid solutions containing at least 4 or more cations or anions, generally with a disordered cation sublattice and an ordered anion sublattice \cite{rost2015entropy}.
These high entropy ceramics include high entropy oxides \cite{rost2015entropy,huang2007microstructure}, carbides \cite{zhou2018high,harrington2019phase}, borides \cite{gild2016high,mayrhofer2018high}, silicides \cite{gild2019high,qin2019high}, sulfides \cite{cui2021high,zhang2018data} and nitrides \cite{chen2004nanostructured,lai2006preparation}, which have many interesting properties due to their increased configurational entropy, such as having high temperature phase stability \cite{lewin2020multi,moskovskikh2020extremely,oses2020high}.
In particular, high entropy metal nitrides (HEMNs) can exhibit much higher hardness and fracture toughness compared to their binary-nitride counterparts \cite{moskovskikh2020extremely,firstov2014thermal,pogrebnjak2018influence}, and along with high melting points and high thermal conductivity \cite{lewin2020multi,pogrebnjak2018influence}, they have many properties that make them ideal for applications such as protective coatings for materials in extreme environments.
The most common defect found in HEMNs are nitrogen vacancies \cite{dippo2020bulk}, which affect the mechanical properties of these materials.
In particular, nitrogen vacancies influence lattice distortion, increase cohesion and formation energy of the HEMN \cite{jiang2023effects}, and decrease hardness \cite{jiang2023effects,dippo2020bulk}.
Therefore, to optimize the mechanical properties of these HEMNs, we need to understand and predict the vacancy formation energy of nitrogen in HEMNs to build more effective HEMN materials.

Nitrogen vacancies and nitrogen content affect the synthesis and mechanical properties of HEMN materials. 
Nitrogen vacancies in binary nitride precursors promote the synthesis of single-phase HEMNs through both enhanced high-temperature diffusion and through the lowering of the enthalpy of mixing \cite{wang2024preparation,wang2024effect,wang2025effect}, creating sub-stoichiometric HEMNs.
Stoichiometry in HEMNs is an important factor that influences these materials' hardness and other mechanical properties; however, the experimental consensus is ambiguous and varies from system to system.
Pogrebnjak \textit{et al.}\ found that the maximum hardness for (TiHfZrVNb)N\textsubscript{x} was found with nitrogen content of 49\% \cite{pogrebnjak2014microstructure}; likewise, Jingchuan Li \textit{et al.}\ found that (MoSiTiVZr)N\textsubscript{x} had a maximum hardness at nitrogen content at 53.7\% \cite{li2022super}, showing that these materials' mechanical properties are best at near stoichiometric ratios.
On the other hand, Bouissil \textit{et al.}\ found that thin films of (TiTaZrHfW)N\textsubscript{x} had the highest hardness when deposited with a nitrogen flow rate of 9\%, which results in a sub-stoichiometric nitride \cite{bouissil2023properties}.
Hang Li \textit{et al.}\ found that increasing the nitrogen ratio in (NbTaMoW)N\textsubscript{x} above x=0.59 lowered the hardness from its maximum value and deteriorated the nitride film \cite{li2021hard}, showing that optimizing nitrogen stoichiometry is important in creating effective HEMN materials.
As nitrogen vacancies are the most common defect found in HEMNs, understanding how the nitrogen vacancies affect the material stoichiometry and mechanical properties is vital in optimizing these materials.

Analyzing how local environment affects anion vacancies in high entropy ceramics \cite{Deml2015,Wexler2021,Witman2023b} and nitrides can be more easily studied through computational methods. 
The vacancy formation energy of all 3$d$--5$d$ binary transition metal nitrides in their stoichiometric B1 crystal structure has been studied using density functional theory (DFT) \cite{balasubramanian2018energetics,lynn2023density}, showing a clear correlation with certain metals having a higher affinity to create nitrogen vacancies than others.
For high entropy ceramics, the effects of local composition have been studied in carbides \cite{zhao2023machine,lu2025coupling} and oxides \cite{chae2022effects}, in which both carbon and oxygen vacancy formation energies correlate strongly with the chemical composition of the nearest-neighbor environment.
Zhao \textit{et al.}\ used a random forest model to predict the vacancy formation energy of carbon vacancies from nearest-neighbor counts up to the fifth shell in (ZrHfNbTa)C \cite{zhao2023machine}.
Lu \textit{et al.}\ used a kernel ridge-regression model to predict carbon vacancy formation energy in (TiZrHfNb)C and a subset of ternary and binary carbides of the same set of constituent metals \cite{lu2025coupling}. 
They found that global features, like the total composition, and local geometric properties like AGNI fingerprints \cite{botu2017machine} and Gaussian symmetry functions \cite{behler2011atom} were not well suited to predicting carbon vacancy formation energy.
On the other hand, they found that features based on the local chemistry were better suited for predicting vacancy formation energy; these features include the nearest-neighbor count as well as the mean and standard deviation of properties based on either the element or binary carbide of each metal in the nearest-neighbor shell, such as atomic weight, electronegativity, and melting point \cite{lu2025coupling}.
For high-entropy oxides, Chae \textit{et al.}\ used a linear regression model with first nearest-neighbor count features to predict the vacancy formation energy of oxygen in (MgCoNiCuZn)O \cite{chae2022effects}.
These papers show that local environment can predict vacancy formation energy in high entropy ceramics; however, these papers focus on a single high-entropy carbide and oxide material each, therefore only studying the correlation between 4--5 metals and the anion vacancy formation energy, while there are a variety of different metals that could be synthesized in high-entropy ceramics.  
Creating computationally inexpensive and predictive models for vacancy formation energy over a multitude of different chemistries is needed to improve how we design and model HEMNs and their mechanical properties. 

Here, we use density-functional theory to calculate the nitrogen vacancy formation energies in 6 different HEMN systems with 8 constituent metals in order to train an effective model. 
A new algorithm creates multiple supercells that maximize the number of nitrogen environments. After relaxing the cells, we use the energy density method \cite{yu2011accurate,yu2011energy} to quantify the energetics of a nitrogen's local environment, and to evaluate its predictive capabilities for vacancy formation energy.
The low, median, and high energy sites are chosen to construct vacancies. With that training set across our 6 different HEMNs, we create a simple linear model to predict nitrogen vacancy formation energy in HEMNs using the local environment, and then evaluate this model on binary and ternary nitrides as well.
Our results show the effectiveness of this model, which agrees with experimental stoichiometry trends, and that the nitrogen vacancy formation energy is primarily determined by the nearest-neighbor metal composition.

\section{Methodology}

\subsection{Enumeration of unique nitrogen environments}

To enumerate the many unique nitrogen environments, we create a set of supercells for a given composition.
With 5 species, and 6 neighbors for a nitrogen atom in the B1 crystal structure, there are $5^6=15625$ possible environments. This large set can be further reduced to 680 symmetrically unique environments. Moreover, this represents 22 possible ``prototype'' environments, corresponding to environments that are chemical permutations of each other.
We use $2\times2\times2$ simple cubic cells with 64 atoms (32 metal and 32 nitrogen) in the B1 crystal structure, with 5 metal elements are distributed over the 32 metal sites; this requires 2 metal appearing 7 times while the other 3 elements appear 6 times. We can make 10 supercells that go through the different possible combinations so that all elements appear equally over the full set of supercells; this is related to the approach used by Kretschmer \textit{et al.}\ \cite{kretschmer2022strain}.
Here, we optimize our 10 supercells to best represent a given HEMN system with the largest possible number of the 680 symmetrically unique environments.
We begin with a single supercell with random chemical assignments; we consider the $5! = 120$ possible permutations of chemistries to generate a set of 120 supercells from which to choose. We incrementally select the supercell that produces the largest increase in the number of unique environments when added to our current set, stopping when we have 10 supercells. We do 5000 different starting supercells to generate sets of ten supercells; we select the set of 10 with the largest number of unique environments.
We use this algorithm instead of the special quasirandom structure (SQS) \cite{zunger1990special}---which is designed to produce a single cell to match the average of an infinite random structure---as we can explicitly determine and maximize the number of nitrogen environments in a set number of supercells, making this method more appropriate for our study.

\begin{figure*}[htbp]
\includegraphics[width=6.4in]{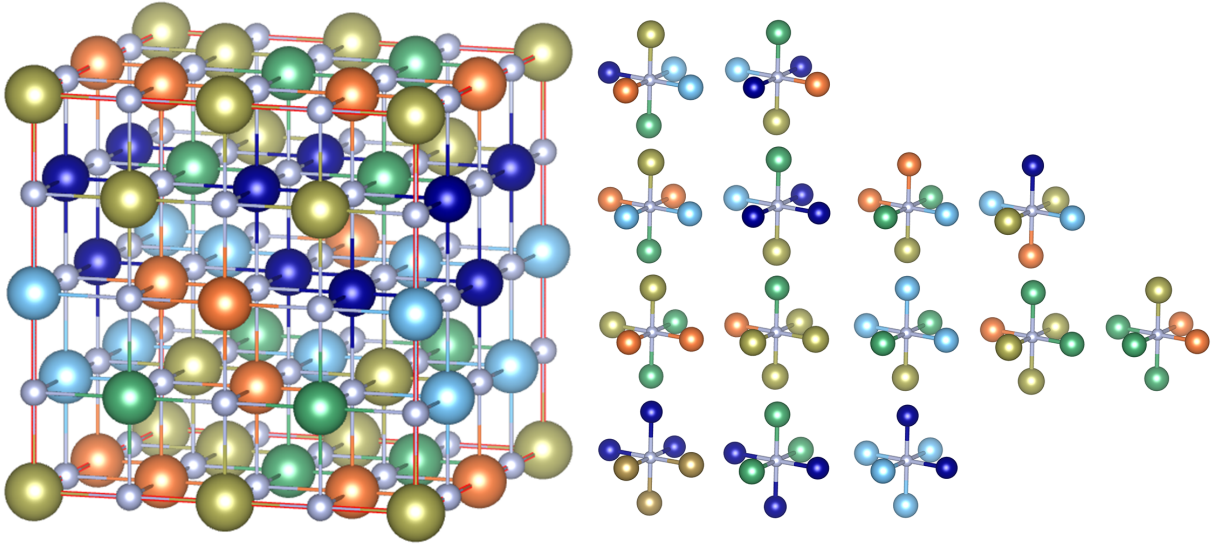}
\caption{(Left) a single B1 high entropy metal nitride (HEMN) supercell used in this study, with the supercell boundary in red; (right) prototype nitrogen environments in this supercell. 
The prototypes are ordered from top to bottom as quinary, quaternary, ternary, and binary environments. 
A nitrogen environment is defined as the octahedron of six metal atoms in the first nearest-neighbor shell of a nitrogen site. 
The prototype environments are A$_3$B$_3$, A\textsubscript{4}B\textsubscript{2}, A\textsubscript{2}B\textsubscript{2}C\textsubscript{2}, A\textsubscript{4}BC, A$_3$B\textsubscript{2}C, A\textsubscript{2}B\textsubscript{2}CD, A$_3$BCD, A\textsubscript{2}BCDE. 
A--E represent the 5 different metal cations contained in each supercell.}
\label{fig:HEMNsupercell}
\end{figure*}

Fig.~\ref{fig:HEMNsupercell}'s 64 atom supercell---with its 14 prototype environments---allow for the sampling of a large fraction of possible nitrogen environments using chemical permutations.
These prototypes geometrically represent every nitrogen environment in our supercells, as each supercell is created through the chemical permutation of the initial supercell.
In total there are 22 possible prototype environments for a hexanary HEMN.
The 8 missing prototype environments consist of a A\textsubscript{6}, A\textsubscript{5}B\textsubscript{1}, A\textsubscript{4}B\textsubscript{2}, A\textsubscript{4}BC, A\textsubscript{2}B\textsubscript{2}C\textsubscript{2}, A\textsubscript{2}B\textsubscript{2}C\textsubscript{2}, A$_3$BCD, and A\textsubscript{2}B\textsubscript{2}CD prototype; A--E represent the 5 metal elements of each HEMN.
Even with 8 missing prototypes, our set of supercells is still effective at sampling nitrogen environments.
Using chemical permutations of the 14 prototypes we have available, we are able to get 310 unique nitrogen environments out of the 320 sites that are accessible with 10 64-atom supercells.
Additionally, out of the 15625 total possible environments for a hexanary HEMN, the 8 missing prototypes represent only 29\%\ of these environments. 
For example, the A\textsubscript{6} environment has only 5 possible chemical permutations, making this environment statistically unlikely and not a priority for our supercell algorithm. Overall, we have 10 supercells for each of the six different HEMNs: (CrHfNbTaTi)N, (CrHfTaTiZr)N, (CrHfNbTiZr)N, (CrNbMoZrV)N, (HfNbTaTiZr)N, and (CrNbTaTiV)N \cite{dippo2020bulk,moskovskikh2020extremely,jin2018mechanochemical}.

\subsection{Supercell relaxation}

To find the lattice constant for each HEMN, we iteratively and simultaneously relax the ten supercells to determine a single equilibrium volume.
The ten supercells in an HEMN each had the same volume, and maintained their cubic geometry to prevent any deviations from the chemistry in a given supercell on the calculations of the EDM and vacancy formation energy.
We start relaxation using an initial lattice constant for each HEMN based on the rule of mixtures with binary nitride lattice constants. We compute the average pressure $\bar P$ across the ten supercells, and then update all volumes from the current volume $V$ using $\Delta V = \bar{P}V/K$; the bulk modulus $K$ is estimated from the average binary nitride bulk moduli. A more accurate guess of $K$ would improve the speed of convergence. We iterate for each HEMN until the average supercell pressure is below 300 MPa (3 kbar).
Having a single equilibrium volume with an average supercell pressure below 300 MPa allows us to calculate properties on all 10 supercells for a HEMN and have the volume and pressure effects be as minimal as possible; additionally, this equilibrium volume gives us the closest estimate to the true lattice constant of the structure with the given chemistries represented in the 10 supercells.

We study the 6 HEMNs (CrHfNbTaTi)N, (CrHfTaTiZr)N, (CrHfNbTiZr)N, (CrNbMoZrV)N, (HfNbTaTiZr)N, and (CrNbTaTiV)N \cite{dippo2020bulk,moskovskikh2020extremely,jin2018mechanochemical}, along with the 8 binary nitrides and 4 ternary nitrides in the B1 crystal structure.
The 8 binary nitrides are the 8 constituent metals in the B1 structure, and the 4 ternary nitrides are (CrNb)N, (CrTi)N, (MoHf)N, and (TiNb)N.
The 6 HEMNs are chosen because they have all been synthesized in bulk in the B1 crystal structure and have near equatomic metal-to-nitrogen ratios.
Each set of HEMN supercells is relaxed until the 300 MPa pressure convergence tolerance is reached; the average pressure for each HEMN is listed in Table~\ref{tab:SC_pressure}.
Some HEMNs have a higher relative $\bar{P}$ than others; however, the volume percent error among all 6 HEMNs is still well below 1\%.
Each binary and ternary nitride is only represented by 1 64-atom supercell; the ternary nitride supercell is created using our algorithm, to maximize the number of symmetrically unique nitrogen environments. 

\begin{table}[ht]
\caption{
Average supercell pressure and expected volume error for each of the 6 HEMNs studied.
$\bar{P}$ is the average supercell pressure for each set of 10 supercells per HEMN, $\sigma$\textsubscript{P} is the standard deviation of the pressures.
The percent error in volume is estimated by dividing $\bar{P}$ by the binary nitride average bulk modulus.
\label{tab:SC_pressure}
}
\begin{ruledtabular}
\begin{tabular}{c c c c}
    HEMN & $\bar{P}$ (MPa) & $\sigma$\textsubscript{P} (MPa) & \% volume error \\
    \hline
    (CrHfNbTiZr)N & 13.4 & 1072 & 0.005 \\
    (CrHfTaTiZr)N & 29.4 & 1206 & 0.011 \\
    (CrHfNbTaTi)N & --15.3 & 1335 & 0.005 \\
    (CrNbMoZrV)N & --252.2 & 1472 & 0.091 \\
    (HfNbTaTiZr)N & --29.2 & 771 & 0.010 \\
    (CrNbTaTiV)N & 0.8 & 1231 & \textless 0.001 \\
\end{tabular}
\end{ruledtabular}
\end{table}

\subsection{Energy density method}

The EDM method creates well-defined atomic energies and volumes to analyze defects in a multitude of solid-state systems. 
The formalism for EDM was first proposed by Chetty and Martin \cite{chetty1992first}, and the version used in this study was implemented for DFT in VASP by Yu \textit{et al.}\ \cite{yu2011energy,yu2011accurate}.
The EDM method rewrites the total DFT energy as an energy density \cite{yu2011energy} which consists of kinetic, exchange-correlation, and classical Coulomb terms, as well as an on-site term corresponding to the integrated energy density within the PAW pseudopotential sphere. 
The EDM energy of an atom is calculated by integrating the energy density over two atom centered volumes, the Bader and charge-neutral volume, and adding the on-site terms.
These two volumes are bound by zero-flux surfaces of the gradient of either the electron density or electrostatic potential, creating basins of attraction which have a maximum at the center of a specific atom. Integrating over these volumes removes the gauge dependence difference between the symmetric and asymmetric forms of the kinetic energy density and classical Coulomb energy density.
The kinetic and exchange-correlation energy density are integrated over the Bader volume, while the classical Coulomb energy density uses the charge-neutral volume.
The energy density, Bader, and charge-neutral volumes are defined on the real-space grid in VASP, with a grid-based weighted integration method to calculate the EDM energies \cite{yu2011accurate}.

We analyze the EDM energy of a nitrogen environment to predict whether or not a nitrogen site is likely to form a vacancy.
The EDM energy of a nitrogen environment $E_\text{env}$ is defined as the sum of the EDM energy of the nitrogen atom $E_\text{N}$ and contributions from each of the 6 nearest-neighbor metal atoms $E_{\text{M}_i}$,
\begin{equation*}
    E_\text{env} = E_\text{N} + \frac{1}{6}\sum_{i=1}^{6} E_{\text{M}_i}.
\end{equation*}
We find that there is nontrivial overlap between the nitrogen atom's Bader volume and the neighboring metal atom's PAW spheres.
The overlap comes from the short metal-nitrogen bond lengths in the HEMNs; the average distance is 2.18\AA. The metal atoms PAW spheres have a mean radius of 1.36\AA, while nitrogen PAW radius is much lower at 0.809\AA.
Additionally, the mean nitrogen Bader volume is 13.2 \AA$^3$, which is larger than the mean Bader volume of every metal atom other than Zr and Mo.
The percentage overlap of a nitrogen atom's Bader volume and an individual nearest-neighbor metal atom's PAW sphere is on average 5\% with the maximum overlap being 18\%.
We see no overlap with nitrogen-nitrogen Bader volumes and PAW spheres.

\subsection{Density functional theory}

\begin{figure}[!htbp]
\includegraphics[width=3.3in]{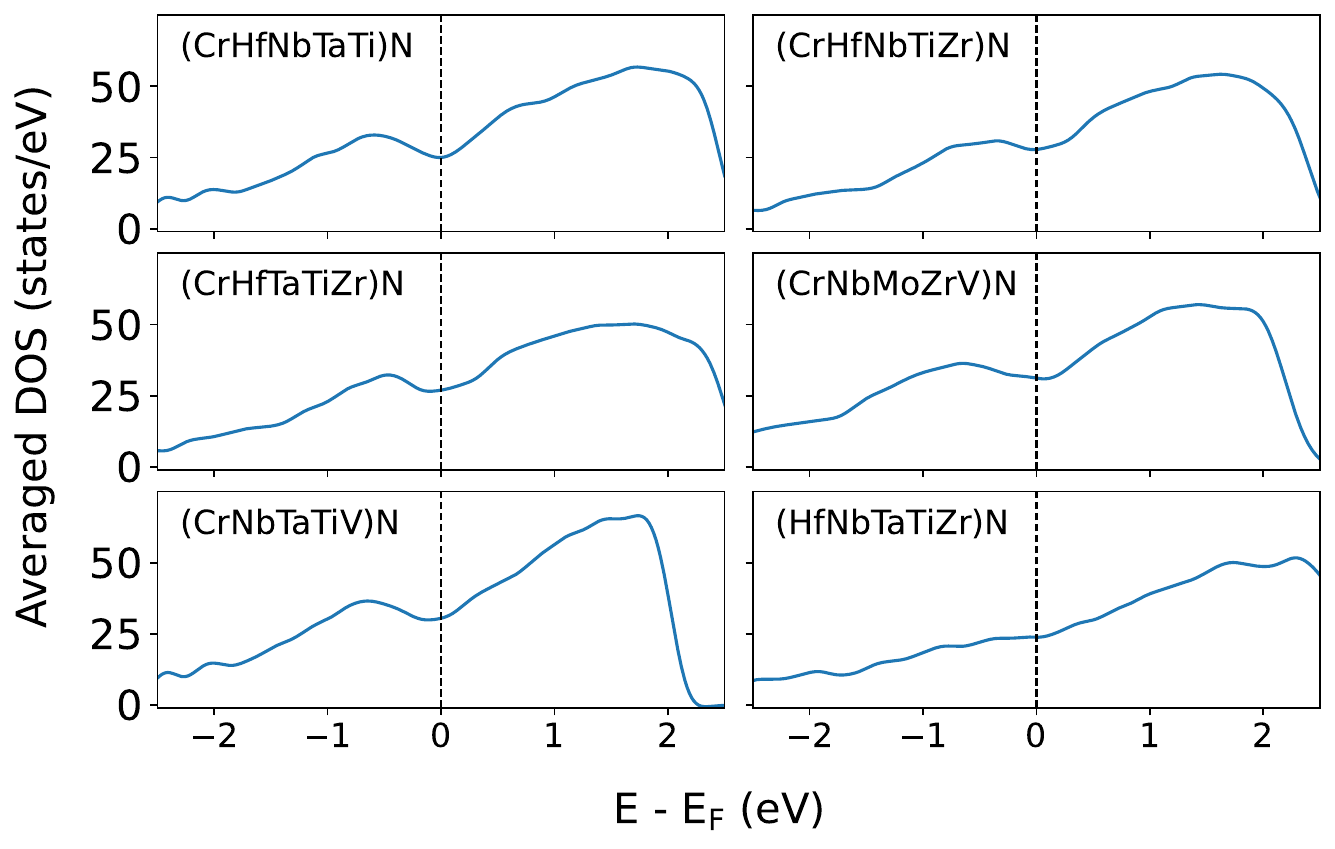}
\caption{
Average density of states (DOS) of all 10 supercells for each of the HEMNs. Energy values are in reference to the average fermi energy, which is represented by the dashed line. The average DOS for each HEMN is taken by interpolating each supercell's DOS onto a common energy spectrum per HEMN, and then averaging the interpolated DOS across all 10 supercells.   
}
\label{fig:DOS}
\end{figure}

The Vienna Ab initio Simulation Package (VASP) \cite{kresse1993ab,kresse1994ab,kresse1996efficiency,kresse1996efficient} computes all supercell energies with modifications to compute the energy density.
We use the projector augmented wave (PAW) method with the Perdew-Burke-Ernzerr (PBE) \cite{perdew1996generalized} formulation of the generalized gradient approximation (GGA) for the exchange-correlation potential. 
A 400 eV plane-wave energy cutoff and a gamma-centered $k$-point grid mesh of $8\times 8 \times 8$ with a order-1 Methfessel-Paxton smearing \cite{methfessel1989high} width of 0.15 eV calculates the DFT forces and energy accurately.
For the DFT geometry relaxations, the force convergence criteria is 5 meV/\AA.
EDM calculations in VASP require a denser real-space grid of $128 \times 128 \times 128$ along with all the same parameters listed above. 
The nitrogen vacancy formation energy $E_\text{vf}$ is given by the DFT energy of the vacancy supercell $E^\text{DFT}_\text{vacancy}$ and the undefected supercell $E^\text{DFT}_\text{supercell}$, along with the nitrogen chemical potential $\mu_{\text{N}}$ as
\begin{equation*}
    E_{\text{vf}} = E^{\text{DFT}}_{\text{vacancy}} - E^{\text{DFT}}_{\text{supercell}} + \mu_\text{N} .
\end{equation*}
The value of $\mu_{\text{N}}$ at zero-temperature is half the DFT energy of a N\textsubscript{2} molecule in a $15 \times 15 \times 15$ \AA\ box: --8.298 eV.
Figure~\ref{fig:DOS} shows the averaged density of states (DOS) for each of the 6 HEMNs studied, which all show non-zero DOS at the Fermi energy, meaning our HEMNs are metals. 
For metallic materials, defects are effectively neutral, so all vacancy formation energy calculations are considered charge neutral. 
We also see that all of our HEMNs have a pseudogap near the Fermi energy, except for (HfNbTaTiZr)N. 
Pseudogaps have also been found in high entropy carbides \cite{lu2025coupling}, indicating this could be an interesting avenue for continued study.
Additionally, DFT calculations have predicted that most transition metal nitrides in the B1 crystal structure are also metallic \cite{lynn2023density}. 

\section{Results}

\begin{figure}[!htbp]
\includegraphics[width=3.3in]{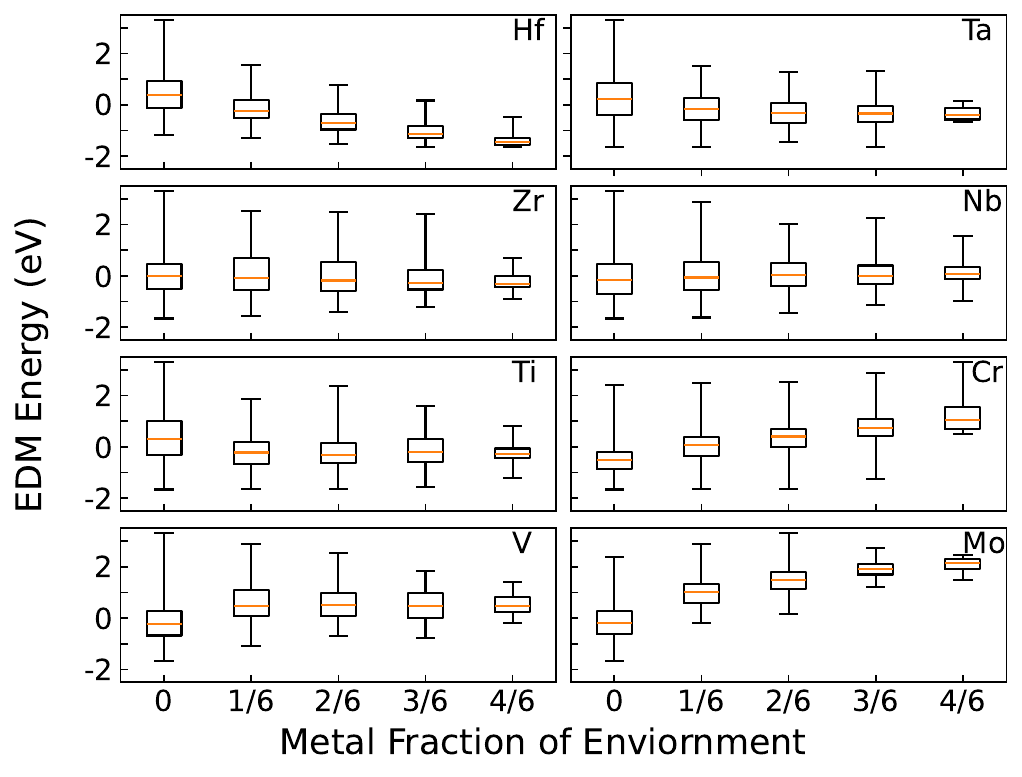}
\caption{
EDM energies of all the HEMN nitrogen sites as a function of the atomic fraction of each metal cation in the site's first nearest-neighbor shell. 
The top and bottom end of the whiskers in the plots represent the maximum and minimum EDM energy, and the orange line is the median.
}
\label{fig:EDM_BOX_WHIS}
\end{figure}

Fig.~\ref{fig:EDM_BOX_WHIS} shows the variation in energy of a nitrogen environment with the local composition, with hafnium decreasing the energy and chromium and molybdenum increasing it.
This correlation between the atomic composition of a nitrogen environment and the EDM energy can be seen in how the median EDM energy changes as atomic composition changes, in Fig.~\ref{fig:EDM_BOX_WHIS} the median EDM energy increases with the atomic fraction of Cr and Mo, while it decreases with more Hf.
The remaining five metals have a weak average influence on the EDM energy.
Nitrogen environments with a higher EDM energy should be less stable and more likely to form a vacancy, while environments with lower EDM energies should be more stable and less likely to form vacancies.
We expect Cr and Mo to lower the vacancy formation energy Hf to raise the vacancy formation energy.

\begin{figure}[!htbp]
\includegraphics[width=3.3in]{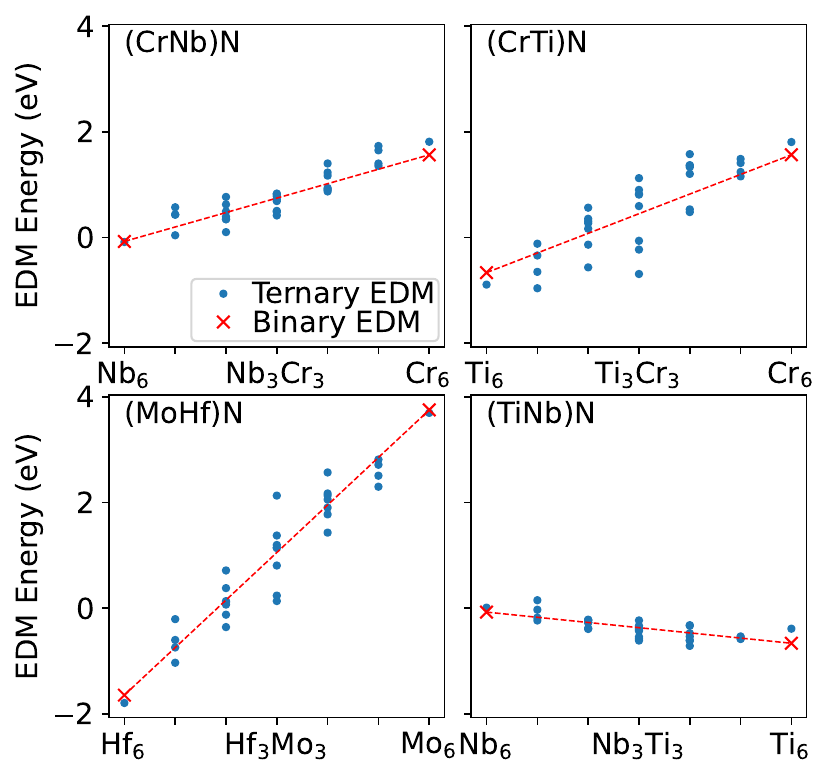}
\caption{
EDM energies of all nitrogen sites in (CrNb)N, (CrTi)N, (MoHf)N, and (TiNb)N as a function of the nearest neighbor composition in the nitride. 
The red endpoints are the EDM energy of the binary nitrides, with a dashed line interpolation.
}
\label{fig:Ternary_EDM}
\end{figure}

Fig.~\ref{fig:Ternary_EDM} shows how the trends in the EDM energy of nitrogen sites in ternary and binary nitrides behave similarly to the trends we found in Fig.~\ref{fig:EDM_BOX_WHIS}.
Ternary nitrides show a stronger correlation between the local composition of a nitrogen environment and its EDM energy than seen in the high entropy nitrides.
For example, in Fig.~\ref{fig:Ternary_EDM}, the change in EDM energy of nitrogen environments in (MoHf)N as the composition changes is much greater than what is seen in the other 3 ternary nitrides as well as in the HEMNs.
Additionally, the EDM energy of binary nitride nitrogen environments is very similar to the EDM energies of the ternary nitride environment with the same nearest-neighbor shell composition, as seen with the red markers in Fig.~\ref{fig:Ternary_EDM}. 
Using these binary nitride EDM energies, we can use a simple linear interpolation to effectively predict ternary nitride EDM energies.
These interpolations are visualized as red dashed lines in Fig.~\ref{fig:Ternary_EDM}, and the MAE between these predictions and the EDM energies of the ternary nitride environments for (CrNb)N, (CrTi)N, (MoHf)N, and (TiNb)N is 0.177, 0.380, 0.298, and 0.098 eV.
The low MAE in these predictions using binary nitride nitrogen EDM energies, along with the correlations we see with both Fig.~\ref{fig:EDM_BOX_WHIS} and Fig.~\ref{fig:Ternary_EDM}, show that the energetics of nitrogen in binary, ternary, and HEMN systems are highly correlated.

Fig.~\ref{fig:Ternary_EDM} shows that the EDM energy of nitrogen environments varies even with the same local composition. In particular, there is large variation in the Mo$_3$Hf$_3$ and Cr$_3$Ti$_3$ environments, with standard deviations of 0.6 eV. 
This variability in EDM energy among environments with the same local composition can be found in all 4 ternary nitrides to varying degrees.
For (CrNb)N, (CrTi)N, (MoHf)N, and (TiNb)N the average standard deviation of EDM energy for compositions with more than 1 site in the dataset is 0.18, 0.36, 0.35, and 0.1 eV respectively. 
Similarly, for the HEMNs, as the mean standard deviation of EDM energy in compositions with more than 1 site in the HEMN dataset is 0.22 eV, and for the ternary nitrides it is 0.25 eV.
This indicates that the EDM energy depends on more than just the first nearest-neighbor composition and that the predictivity of the local composition on the site's EDM energy has a RMSE floor of 0.22 eV.

\begin{table*}[!htbp]
\caption{EDM and vacancy formation energy ($E_\text{vf}$) values for a sub-set of nitrogen sites for (CrNbMoZrV)N, (CrNbTaTiV)N and (HfNbTaTiZr)N, which are the 3 HEMNs with the lowest average $E_\text{vf}$. 
The sub-set of sites consists of 5 nitrogen vacancies with the low EDM energies, 5 sites with high EDM energy, and 3 median EDM energy sites. 
EDM and $E_\text{vf}$ values are in eV. 
The nearest-neighbors are represented as counts of the cations in the first nearest-neighbor shell of the vacancy site. 
Here we see a common trend that sites with low EDM energies have high $E_\text{vf}$ and sites with high EDM energies have low $E_\text{vf}$.
We also see correlation with the metal cations in the vacancy environment and the value of the $E_\text{vf}$. 
For example, sites with more Cr, Mo, and Nb in their environment will tend to have lower $E_\text{vf}$.
\label{tab:low_Evf_table}
}
\begin{ruledtabular}
\begin{tabular}{c c c c c c c c c}
    \multicolumn{3}{c}{(CrNbMoZrV)N} & \multicolumn{3}{c}{(CrNbTaTiV)N} & \multicolumn{3}{c}{(HfNbTaTiZr)N} \\
    EDM & $E_\text{vf}$ & nearest-neighbor & EDM & $E_\text{vf}$ & nearest-neighbor & EDM & $E_\text{vf}$ & nearest-neighbor \\
    \hline \\
    --0.677 & 2.253 & Zr$_2$VNb$_2$Cr & --1.174 & 1.584 & TiTa$_3$NbCr & --1.657 & 2.271 & Hf$_3$Ta$_3$ \\
    --0.437 & 2.335 & Zr$_3$V$_2$Cr & --1.075 & 1.220 & TiVTa$_2$Cr$_2$ & --1.617 & 2.679 & Hf\textsubscript{4}TaNb \\
    --0.326 & 2.179 & Zr$_2$VNb$_3$ & --0.999 & 2.064 & Ti$_3$VTaNb & --1.611 & 3.060 & Hf\textsubscript{4}Ta$_2$ \\
    --0.278 & 1.841 & ZrV$_2$Nb$_3$ & --0.906 & 1.987 & Ti$_2$Ta$_3$Nb & --1.552 & 3.220 & Hf\textsubscript{4}ZrTi \\
    --0.257 & 1.487 & V$_2$Nb$_3$Cr & --0.893 & 1.561 & TiVTa$_2$Nb$_2$ & --1.436 & 2.914 & Hf$_3$Ta$_3$ \\
    \\
    \\
    1.186 & 1.029 & V$_2$Nb$_2$CrMo & 0.126 & 1.488 & TiVTa$_2$NbCr & --0.560 & 1.336 & HfZrTi$_2$TaNb \\
    1.120 & 1.220 & V$_3$CrMo$_2$ & 0.114 & 1.472 & TiVTa$_2$NbCr & --0.546 & 2.299 & Zr$_3$Ti$_3$ \\
    1.089 & 1.518 & ZrV$_2$NbCrMo & 0.063 & 1.376 & TiVTa$_2$Cr$_2$ & --0.427 & 2.703 & HfZr$_2$TiTaNb \\
    \\
    \\
    2.488 & 1.074 & Zr$_2$CrMo$_3$ & 0.952 & 0.942 & TiTaNbCr$_3$ & 0.087 & 0.935 & TiTa$_2$Nb$_3$ \\
    2.543 & 1.356 & ZrV$_2$Cr$_2$Mo & 1.070 & 1.367 & V$_3$Cr$_3$ & 0.106 & 1.435 & Ti$_2$TaNb$_3$\\
    2.732 & 0.514 & Cr$_3$Mo$_3$ & 1.076 & 0.997 & TiVNbCr$_3$ & 0.153 & 0.918 & TiTa$_3$Nb$_2$ \\
    2.884 & 0.889 & VNbCr$_3$Mo & 1.085 & 0.473 & VTaNb$_2$Cr$_2$ & 0.165 & 0.568 & Ta$_2$Nb\textsubscript{4} \\
    3.297 & 0.675 & Cr\textsubscript{4}Mo$_2$ & 1.526 & 0.807 & TaNb$_2$Cr$_3$ & 0.228 & 1.665 & Ta$_3$Nb$_3$ \\
\end{tabular}
\end{ruledtabular}
\end{table*}

\begin{table*}[!htbp]
\caption{EDM and vacancy formation energy ($E_\text{vf}$) values for a subset of nitrogen sites for (CrHfNbTaTi)N, (CrHfNbTiZr)N and (CrHfTaTiZr)N, which are the 3 HEMNs with the highest average $E_\text{vf}$. 
Sites with more Hf, Zr, and Ti tend to have the highest $E_\text{vf}$. Similar to Table~\ref{tab:low_Evf_table}, we see more Cr and Nb in the sites with the lowest $E_\text{vf}$; however, there is not as high of a concentration of these elements in the vacancy environemnts which may have resulted in the higher average $E_\text{vf}$ of these HEMNs compared to the HEMNs in Table~\ref{tab:low_Evf_table}.
\label{tab:high_Evf_table}
}
\begin{ruledtabular}
\begin{tabular}{c c c c c c c c c}
    \multicolumn{3}{c}{(CrHfNbTaTi)N} & \multicolumn{3}{c}{(CrHfNbTiZr)N} & \multicolumn{3}{c}{(CrHfTaTiZr)N} \\
    EDM & $E_\text{vf}$ & nearest-neighbor & EDM & $E_\text{vf}$ & nearest-neighbor & EDM & $E_\text{vf}$ & nearest-neighbor \\
    \hline \\
    --1.624 & 2.798 & Hf$_3$TiCr & --1.614 & 2.947 & Hf$_3$Ti$_2$Nb & --1.635 & 2.711 & Hf\textsubscript{4}Cr$_2$ \\
    --1.574 & 2.743 & Hf$_3$TiTaCr & --1.578 & 3.034 & Hf$_3$ZrTi$_2$ & --1.633 & 2.928 & Hf$_3$Ti$_2$Ta \\
    --1.490 & 2.949 & Hf$_3$Ti$_2$Ta & --1.556 & 3.080 & Hf$_3$ZrCr & --1.545 & 2.861 & Hf$_3$ZrTiTa \\
    --1.458 & 2.625 & Hf$_3$TiTaNb & --1.497 & 2.691 & Hf$_3$Ti$_2$Cr & --1.516 & 1.917 & Hf$_2$Ta$_3$Cr \\
    --1.398 & 2.380 & Hf$_2$Ti$_2$TaCr & --1.489 & 2.853 & Hf$_3$ZrTiCr & --1.496 & 2.903 & Hf$_3$ZrTiCr \\
    \\
    \\
    --0.251 & 1.196 & HfTa$_2$Nb$_3$ & --0.235 & 1.723 & Ti$_2$Nb\textsubscript{4} & --0.203 & 2.562 & Hf$_2$Zr$_2$TaCr \\
    --0.216 & 1.833 & Hf$_2$Ta$_2$NbCr& --0.165 & 2.155 & Hf$_2$TiNb$_2$Cr & --0.198 & 2.561 & Hf$_2$ZrTiTaCr \\
    --0.206 & 2.202 & HfTi$_2$TaNbCr & --0.148 & 2.175 & Zr$_3$TiNbCr & --0.173 & 2.224 & HfZrTiTa$_2$Cr \\
    \\
    \\
    1.179 & 1.654 & Ti$_2$Ta$_2$Cr$_2$ & 1.582 & 1.907 & Ti$_3$NbCr$_2$ & 1.308 & 1.000 & Ta$_3$Cr$_3$ \\
    1.184 & 1.169 & TiTaNbCr$_3$ & 1.884 & 1.349 & TiNb$_2$Cr$_3$ & 1.311 & 1.605 & Ti$_3$Cr$_3$ \\
    1.256 & 1.327 & Nb$_2$Cr\textsubscript{4} & 2.014 & 1.109 & ZrNb$_2$Cr$_3$ & 1.350 & 2.150 & Zr$_2$Ti$_2$Cr$_2$ \\
    1.581 & 1.307 & Ti$_2$NbCr$_3$ & 2.269 & 1.014 & Nb$_3$Cr$_3$ & 1.438 & 1.874 & Zr$_3$Cr$_3$ \\
    1.765 & 0.793 & Nb$_3$Cr$_3$ & 2.390 & 1.480 & Ti$_2$Cr\textsubscript{4} & 1.956 & 1.655 & Ti$_2$Cr\textsubscript{4} \\   
\end{tabular}    
\end{ruledtabular}
\end{table*}    

Table~\ref{tab:low_Evf_table} and Table~\ref{tab:high_Evf_table} show for the 6 HEMNs, the nitrogen sites with the highest and lowest EDM energy have similar metals in their respective environments, which directly affect the site's vacancy formation energy.
For each of the HEMNs, we select 13 nitrogen sites to construct vacancies: 5 nitrogen sites with the highest EDM energy and lowest EDM energy, and 3 median EDM energy sites. 
We select this subset of 320 available sites in each HEMN to determine the strength of correlation between EDM and the vacancy formation energy.
Table~\ref{tab:low_Evf_table} shows the EDM, vacancy formation energy, and local environment data for the 3 HEMNs with the lowest average vacancy formation energy, while Table~\ref{tab:high_Evf_table} shows the data from the 3 HEMNs with the highest average vacancy formation energy.
Within a given HEMN, sites with high/low EDM energy do generally have lower/higher vacancy formation energy; however, the EDM energy does not necessarily predict whether one site will have a higher or lower vacancy formation energy than another.
For example, in Table~\ref{tab:low_Evf_table} for (CrNbMoZrV)N, the site with the lowest EDM energy does not have the highest vacancy formation energy, the site with the highest vacancy formation energy has the second-lowest EDM energy. 
We also note that in Table~\ref{tab:high_Evf_table}, Hf is commonly featured in the high vacancy formation energy section in for 3 HEMNs, while there is no Hf represented at all in the low vacancy formation energy environments.

\begin{figure}[htbp]
\includegraphics[width=3.3in]{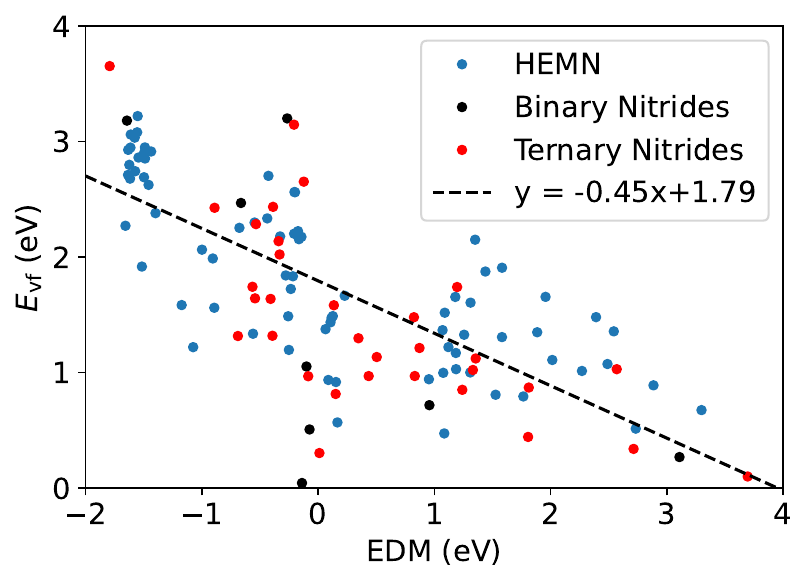}
\caption{
EDM and vacancy formation energies ($E_\text{vf}$) for HEMN, ternary, and binary nitride nitrogen sites.
A linear fit (dashed line) of the EDM values to the $E_\text{vf}$ values shows a negative correlation with an $R^2=0.534$.
}
\label{fig:EDM_v_Evf}
\end{figure}

Fig.~\ref{fig:EDM_v_Evf} shows the correlation between EDM energy of nitrogen sites and vacancy formation energy.
The negative correlation between EDM and vacancy formation energy is as expected; however, as seen by the $R^{2}$ value of 0.534, EDM can only qualitatively predict if a nitrogen site will have a high/low vacancy formation energy.
In particular, we see with nitrogen sites with EDM energies around 0 eV there is a considerable range of vacancy formation energies, which shows that the EDM prediction of vacancy formation energy works best for sites with low/high vacancy formation energy.
Binary nitride EDM energies in particular do not correlate as well with the vacancy formation energy, as many of the binary nitrides have EDM energies near 0 eV.
It is also worth noting the range of EDM values is larger than the spread in vacancy formation energies.
The lack of strong predictive power for the EDM energies is worth investigating further, along with developing models to predict the vacancy formation energy.

\begin{table}[!htbp]
\caption{
Coefficients for each metal cation from the nn-model trained on HEMN DFT calculated nitrogen vacancy formation energy, $E_\text{vf}$.
The coefficients minimizes the mean squared error, weighted by the inverse of the HEMN $E_\text{vf}$.
Multiplying the coefficients by 6 gives the predicted binary nitride $E_\text{vf}$, which is similar to the DFT calculated binary nitride $E_\text{vf}$
All coefficients and $E_\text{vf}$ values are in units of eV.
\label{tab:Model_coefs}
}
\begin{ruledtabular}
\begin{tabular}{c c c c}
     \shortstack{Metal \\ Cation $\alpha$} & \shortstack{$E_{\text{vf},\alpha}$ model \\ coefficients} & \shortstack{predicted \\ binary nitride $E_\text{vf}$}& \shortstack{true \\ binary nitride $E_\text{vf}$}\\
    \hline 
    Hf & 0.600 & 3.600 & 3.181 \\
    Zr & 0.517 & 3.102 & 3.200 \\
    Ti & 0.455 & 2.730 & 2.468 \\
    V & 0.330 & 1.980 & 1.052 \\
    Cr & 0.143 & 0.858 & 0.718 \\
    Ta & 0.139 & 0.834 & 0.044 \\
    Nb & 0.125 & 0.750 & 0.507 \\
    Mo & 0.015 & 0.090 & 0.268 \\
\end{tabular}
\end{ruledtabular}
\end{table}

We model the vacancy formation energy by summing contributions from the six nearest neighbor metal atoms, with coefficients in Table~\ref{tab:Model_coefs}.
This nn-model has a vacancy energy contribution $E_{\text{vf}},\alpha$ for each species $\alpha$ in the first neighbor shell,
\begin{equation*}
E_{\text{vf}} = \sum^{6}_{i=1} E_{\text{vf},\alpha_{i}}.
\label{eqn:model}    
\end{equation*}
This is equivalent to counting the number of neighbors for each species. To find the parameters, we minimize the mean squared error weighted by the inverse of the DFT vacancy formation energies to prioritize low vacancy formation energy sites, using all 78 calculated nitrogen vacancies, with 8 different metal neighbors.
The nn-model was chosen over a model that includes the nitrogen EDM energies in the feature set since including the EDM energies only improved the mean-squared error for the HEMNs by less than 0.01 eV.
The nn-model provides a good prediction of the HEMN vacancy formation energies with a mean-absolute error of 0.149 eV while also being easily interpretable over more sophisticated models.
As predicted by the correlation between EDM and each metal in a nitrogen environment in Fig.~\ref{fig:EDM_BOX_WHIS}, we see the coefficients for Hf and Mo show that more Hf in a nitrogen environment will result in a higher vacancy formation energy, while more Mo will result in lower vacancy formation energies. 
Additionally, the coefficients from our nn-model also show a qualitative predictive ability for the binary nitride vacancy formation energy as seen in Table~\ref{tab:Model_coefs}, showing how the effects of these 8 metals on HEMN vacancy formation energies extend to binary and ternary nitrides as seen with the EDM energies in Fig.~\ref{fig:Ternary_EDM}.

\begin{figure*}[htbp]
  \includegraphics[width=3.2in]{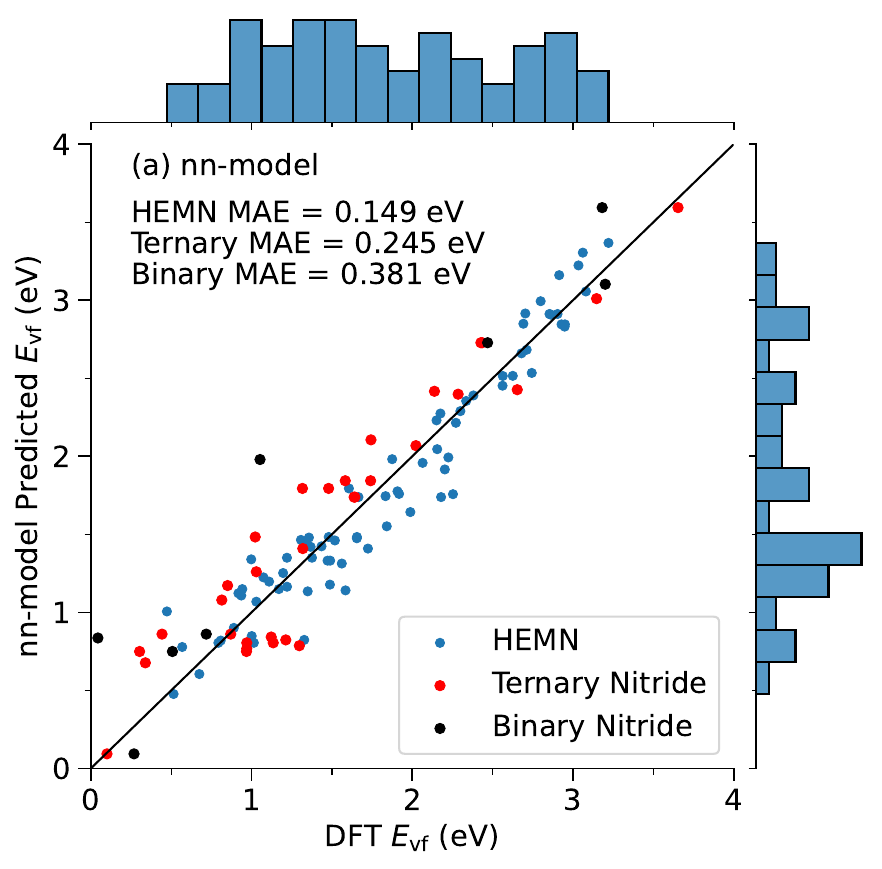}
  \includegraphics[width=3.2in]{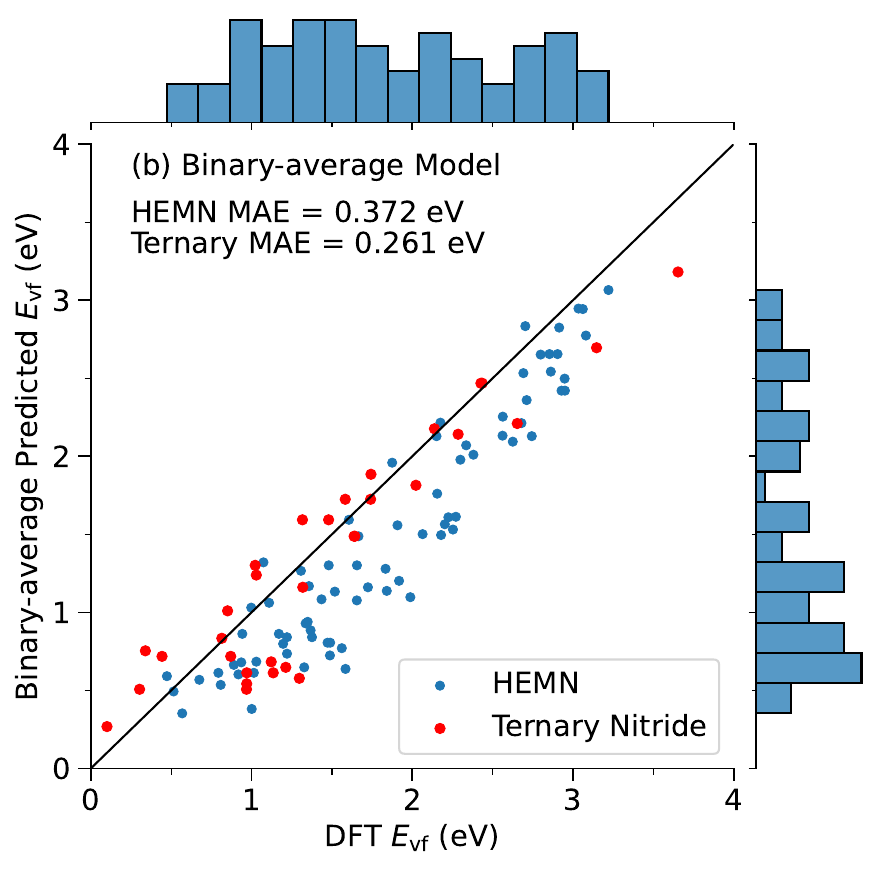}
\caption{Parity plots of results from our nn-model (left) trained on HEMN nitrogen-vacancy environments, versus our Binary-average $E_\text{vf}$ model (right) for ternary and HEMNs created by averaging DFT calculated binary-nitride $E_\text{vf}$ values of the constituent metals in ternary and HEMN vacancy environments.
The nn-model's feature set is the counts of the metal atoms in the first nearest-neighbor shells. 
The histograms represent the range of values for the model-predicted and DFT calculated HEMN $E_\text{vf}$.}  
\label{fig:Parity_plots}
\end{figure*}

Fig~\ref{fig:Parity_plots} shows that the nn-model built on HEMN data predicts HEMN vacancy formation energies well, but performs poorly for ternary and binary systems, and vice versa for models built with binary nitride vacancy formation energy data.
The vacancy formation energies for the ternary nitrides consist of 8 sites per ternary nitride based on the following nearest-neighbor compositions : 1 A$_6$, 1 A$_5$B, 1 A$_4$B$_2$, 2 A$_3$B$_3$, 1 A$_2$B$_4$, 1 AB$_5$, and 1 B$_6$. 
The nn-model, which is only trained on HEMN vacancy formation energies, provides a good prediction for HEMNs with a MAE of 0.149 eV, and is less predictive for ternary nitride $E_\text{vf}$ values with a MAE of 0.245 eV, and MAE of 0.381 eV for the binary nitrides.
The HEMN vacancy formation energy dataset does not contain any environments that contain A$_5$B or A$_6$ compositions as seen in Fig.~\ref{fig:HEMNsupercell}.
Training the nn-model on only HEMN data and not having A$_5$B or A$_6$ environments in the training set would naturally result in the model not performing as well on ternary and binary nitrides; however, given the MAE of 0.245 and 0.381 eV for ternaries and binaries we can see that effect of these various metals in the 1nn shell around a nitrogen site applies to all 3 nitride systems we have studied. In addition, there are some anomalies in the binary nitride data (see Table~\ref{tab:Model_coefs}); TaN, NbN, and MoN vacancy energies are for the B1 structure despite that structure not being the lowest energy nitride. And as a HEMN rarely samples such environments, there is significant disagreements between the binary and HEMNs.

To investigate the effect that the different types of nitrogen environments play on the prediction of nitrogen vacancy formation energy we used a binary nitride based model to predict vacancy formation energies based on binary nitride vacancy formation energies \cite{lim2023predicting}. 
This model in Fig.~\ref{fig:Parity_plots} predicts the vacancy formation energy of a nitrogen site by averaging the binary-nitride vacancy formation energies of each of the 6 nearest-neighbor metals in a nitrogen environment.
This alternate model predicts ternary nitrides vacancy formation energies with a similar MAE of 0.261, but gives a MAE of 0.372 for the HEMNs.
Additionally, the binary-average model underestimates the DFT vacancy formation energies for both ternary and HEMNs, highlighting a more systematic issue with using only binary nitride vacancy formation energies to predict more disordered nitride systems.
The results seen in Fig.~\ref{fig:Parity_plots} show how vacancy formation energies in HEMN, ternary, and binary nitrides can be predicted effectively with relatively simple models, but also highlights the importance of correctly choosing a training dataset that is representative of the types of nitrogen environments that are being predicted.

Across our EDM and vacancy formation energy results, we find a common trend in which more Hf, Zr, Ti, and V in a nitrogen site's environment increases the vacancy formation energy, while more Ta, Nb, Cr, and Mo decreases the vacancy formation energy.
The correlation between these sets of metals and whether a HEMN nitrogen site will have high or low vacancy formation energy can be understood through the vacancy formation energy and phase stability of the corresponding binary nitrides.
For HfN, ZrN, TiN, and VN, the most stable phase is the B1 rocksalt structure, which is the crystal structure which the HEMNs are simulated in \cite{balasubramanian2018energetics}.
The metals that correspond to these binary nitrides also form stable environments in our HEMNs as they have large coefficients in our nn-model.
However, this is not the case for TaN, NbN, CrN, and MoN.
For TaN, NbN and MoN, recent studies have found that hexagonal phases are more stable than the cubic B1 structure \cite{grumski2013ab,kanoun2007structure,hart2000phonon,zou2016discovery,zhao2015phase}.
Where the ground state of TaN is the P$\overline{6}$2m structure \cite{grumski2013ab}, NbN is the P$\overline{6}$m2 structure \cite{zhao2015phase}, and MoN is in the Wurtzite P$6_{3}$mc structure \cite{kanoun2007structure}. 
Additionally, in CrN the anti-ferromagnetic B1 structure is the ground state structure, and the non-magnetic B1 CrN is unstable \cite{zhou2014structural,mavromaras1994investigation}.
All our HEMNs were simulated in the non-magnetic B1 crystal structure; as a result, the metals that correspond to binary nitrides that are not stable in either the B1 structure (TaN, NbN, and MoN) or the non-magnetic case (CrN) would be less stable in our HEMNs.
We can see this result in the coefficients from our nn-model in Table~\ref{tab:Model_coefs}, where Ta, Nb, Mo, and Cr not only have the smallest coefficients, but their binary nitrides have the lowest vacancy formation energies in the B1 structure.   
These results, along with our binary-average model in Fig.~\ref{fig:Parity_plots}, show how data from binary nitrides can be used to understand and predict properties like vacancy formation energy in HEMNs.

\begin{figure}[htbp!]
\centering
\includegraphics[width=3.3in]{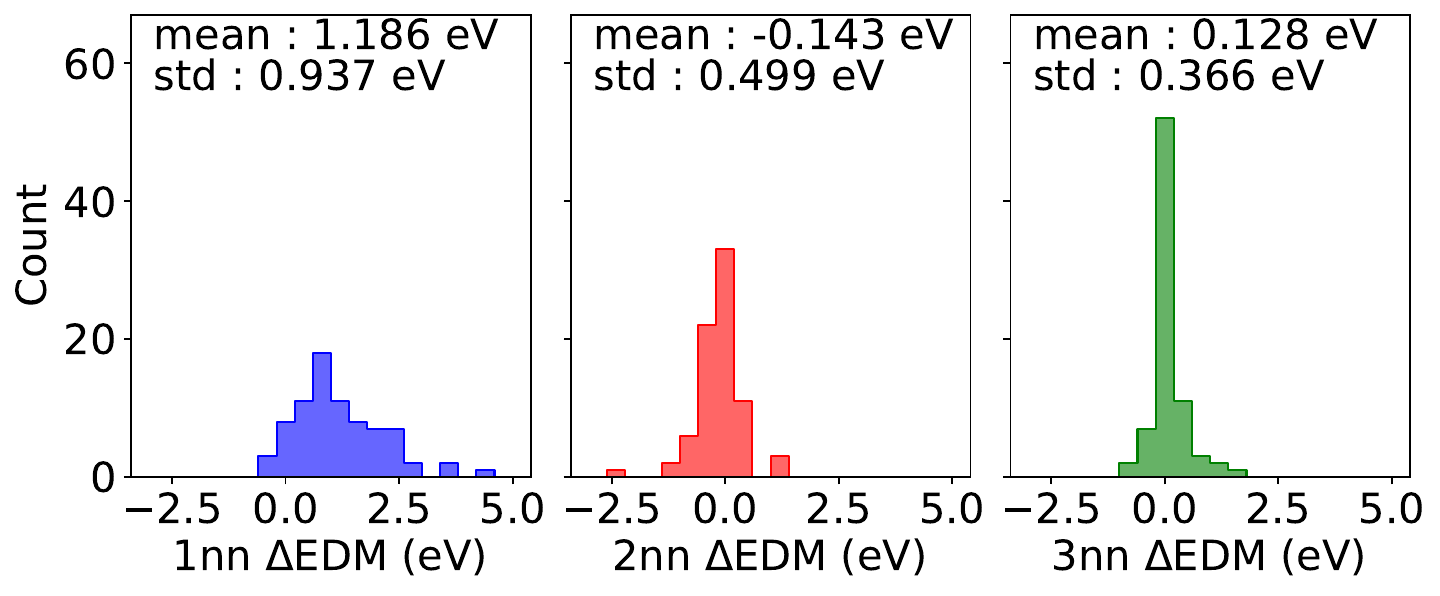}
\vspace{2em}
\includegraphics[width=3.3in]{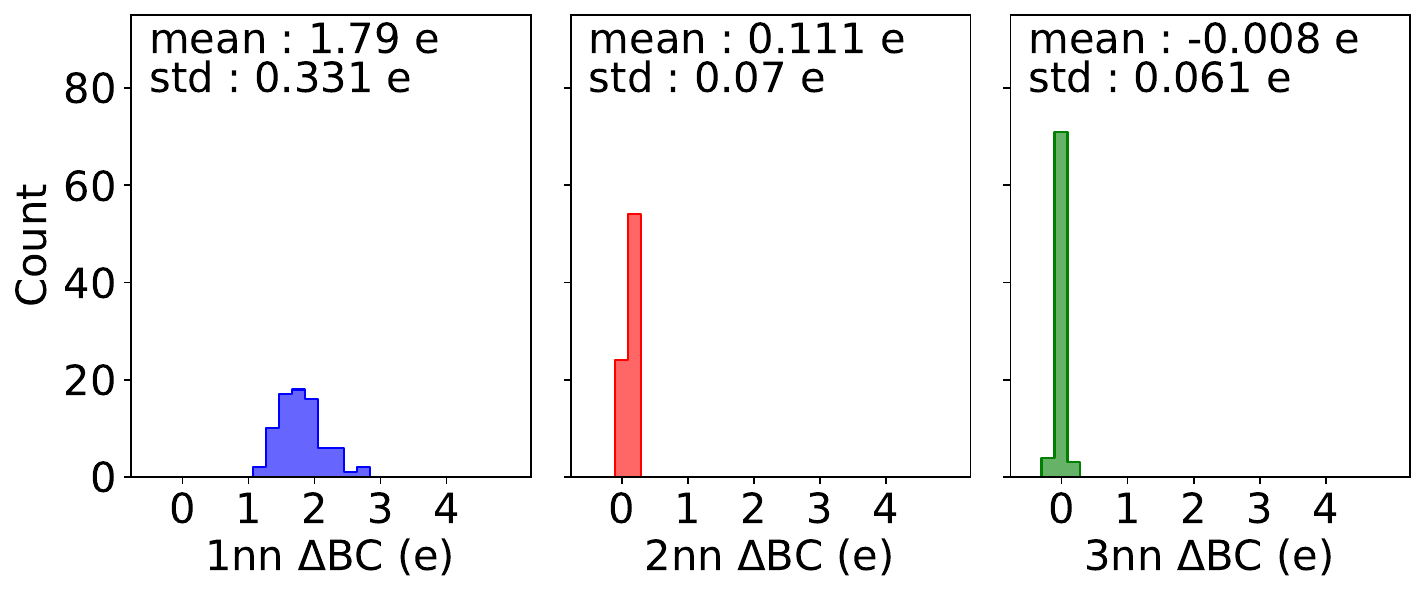}
\caption{
Change in EDM energy ($\Delta$EDM) and Bader charge ($\Delta$BC) of nearest-neighbor atoms surrounding all nitrogen vacancy sites after the nitrogen atom is removed and atoms are re-relaxed.
The change in EDM energy and BC of a nearest-neighbor shell is defined as the sum of the change in EDM energies or BC of each atom in each nearest-neighbor shell.
}
\label{fig:Delta_EDM_BC}
\end{figure}

Fig.~\ref{fig:Delta_EDM_BC} shows how the EDM energy and Bader charge changes for the first three nearest-neighbor shell atoms surrounding a nitrogen site after a vacancy is created, indicating that energy and charge differences in vacancy formation are highly localized.
The change in EDM energy ($\Delta \text{EDM}$) of 1nn shell metals surrounding a vacancy is much greater than what is found in the 2nn and 3nn shell atoms, as seen by the mean and standard deviation of the $\Delta \text{EDM}$ in each nn shell.
The nitrogen environment EDM energy only tells us information about the energetics of the nitrogen site pre-vacancy, so this relatively large change in the 1nn shell $\Delta \text{EDM}$ on the order of 1 eV highlights the limits of using the nitrogen environment EDM energy as a predictor for vacancy formation energies.
Additionally, on average, the formation of a nitrogen vacancy increases the EDM energy of the nn metals.
This increase in EDM energy is expected, as this means the nn metals are less stable, which is to be expected with the creation of a vacancy in the HEMN.
In regard to the relatively small $\Delta \text{EDM}$ in the 2nn and 3nn shell atoms, this explains why our nn-model can predict vacancy formation energies in HEMNs.
The mean change in $\Delta \text{EDM}$ for the 2nn and 3nn shell is --0.143 eV and 0.128 eV, suggesting that the outer neighbor shell contributions are both smaller and cancel on average.
Therefore, having a model which excludes these atoms will not affect the model as heavily as just the 1nn shell atoms. 
Additionally, in Fig.~\ref{fig:Delta_EDM_BC}, we see how the effects of creating a nitrogen vacancy on the Bader charge of the atoms in our HEMNs is also localized to the first nearest-neighbor shell. 
We can see that the mean change in Bader charge ($\Delta$BC) for the 1nn shell atoms is significant at 1.79 e, with a maximum change in Bader charge of 2.7 e. 
In comparison, we can see that the change in Bader charge for the 2nn and 3nn shell atoms is insignificant.
Overall, the vacancy formation energy is highly localized, which can guide what features and models should be used to predict these types of defects in various nitride systems. 

\begin{figure*}[htbp]
    \includegraphics[width=3.2in]{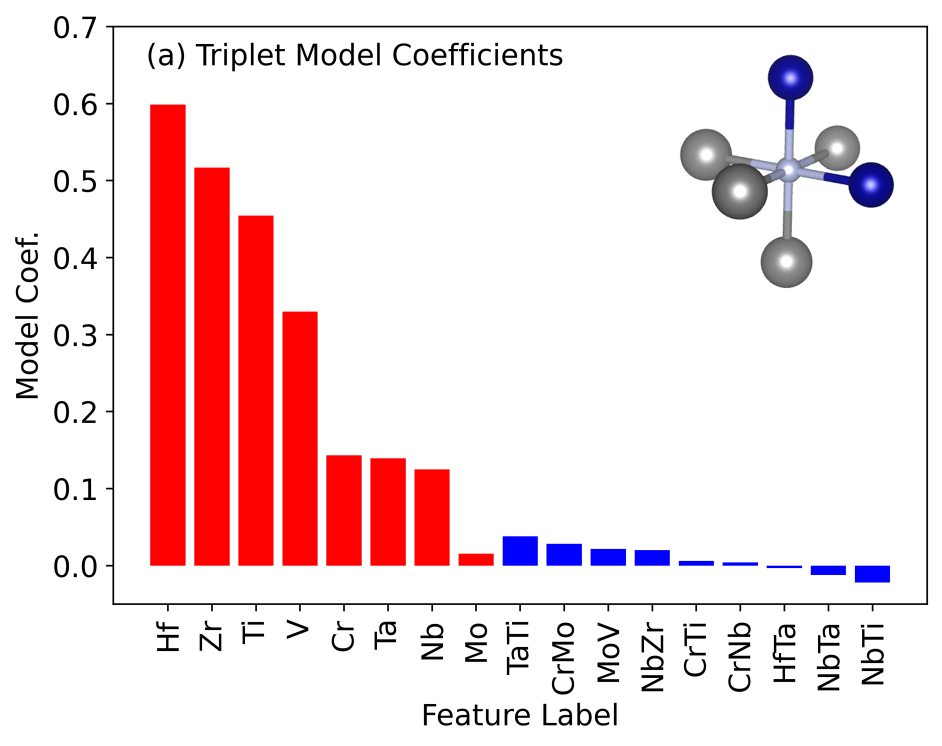}
    \includegraphics[width=3.2in]{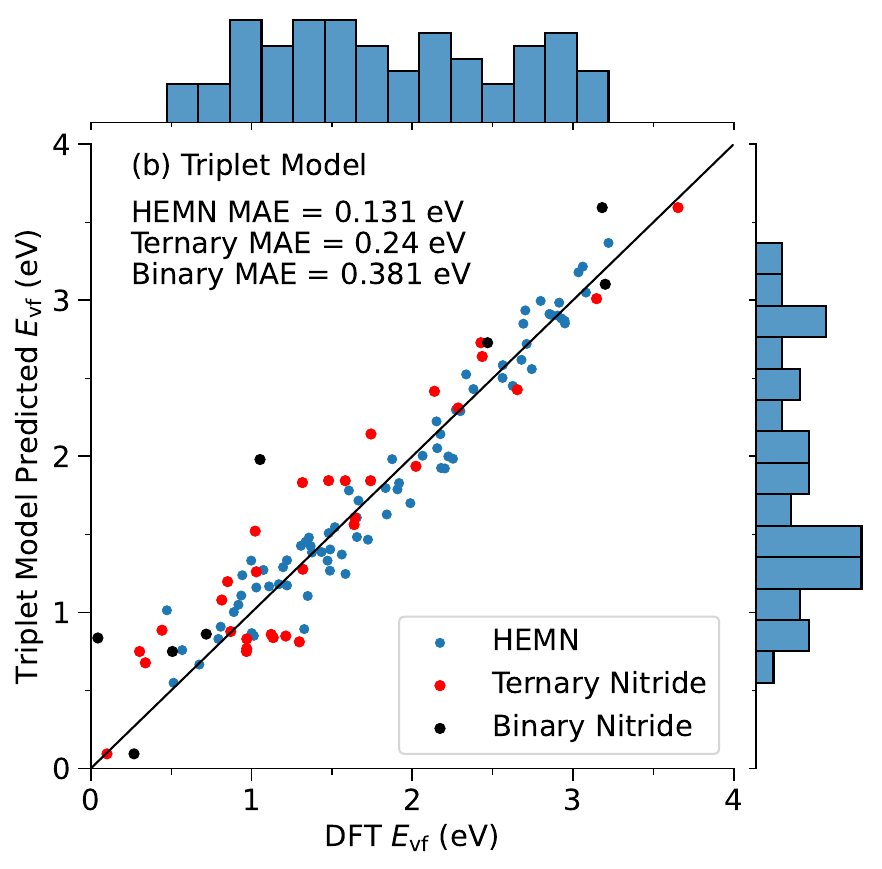}
\caption{
Triplet model coefficients with original nn-model coefficients (red) and triplet correction coefficients (blue) from training a LASSO model on the residuals of the nn-model, with corresponding parity plot.
The blue atoms in the inset image geometrically represent the metals that form a triplet in a nitrogen environment.
The triplet model shows slight improvement over nn-model in MAE of HEMN and ternary nitride $E_\text{vf}$. 
}  
\label{fig:Triplet_Model}
\end{figure*}

Fig.~\ref{fig:Triplet_Model} shows the results of a model that includes corrections to our nn-model by including features based on triplet clusters that are found in our nitrogen environments.
The triplets are defined as groups of three atoms in a nitrogen environment: two metal and one nitrogen. 
Our triplet features exclude triplets in which the two metals are directly opposite each other from the nitrogen site; each nitrogen environment has 12 triplets. The triplets can provide additional geometric information on the nitrogen environment in addition to the features used in our nn-model. 
We create a new model, which we will call the triplet model, where we train a LASSO model on the residuals of our NN model. 
Using LASSO on the residuals provides a feature reduction to avoid overfitting; the regularization parameter was chosen through cross-validation to be 0.017.
There are 36 triplet features which are reduced to 9 features which provide the greatest reduction in RMSE compared to our nn-model, these 9 features and their coefficients are seen in Fig.~\ref{fig:Triplet_Model}.
As seen in the parity plot, our triplet model reduces the MAE of our HEMN predictions by 0.018 eV, with negligible effects on the ternary and binary nitrides. 
The small effect on the ternary and binary nitride predictions is expected as these 9 triplet features do not impact any binary nitride system, and only the CrNb and NbTi features impact (CrNb)N and (TiNb)N, respectively. 
Additionally, the coefficients of the LASSO features can be both positive and negative, which suggests bias in nn-model predictions of vacancy formation energies for environments with these features. 
The TaTi, CrMo, MoV, NbZr, CrTi, and CrNb triplet features suggest underestimation by our nn-model since their coefficients are positive, while HfTa, NbTa, and NbTi features suggest overestimation by our nn-model.
The relatively small difference in MAE in the HEMN predictions shows us that the nearest-neighbor count features in our nn-model are the primary nitrogen environment features that impact the prediction of the vacancy formation energy, and that more complex chemistry-based features can only provide small corrections.
Therefore, by using our nn-model and triplet model, we can see that we are reaching the limit of how well we can predict the vacancy formation energies from just the local environment.

\section{Discussion}

\begin{figure}[htbp!]
\includegraphics[width=3.3in]{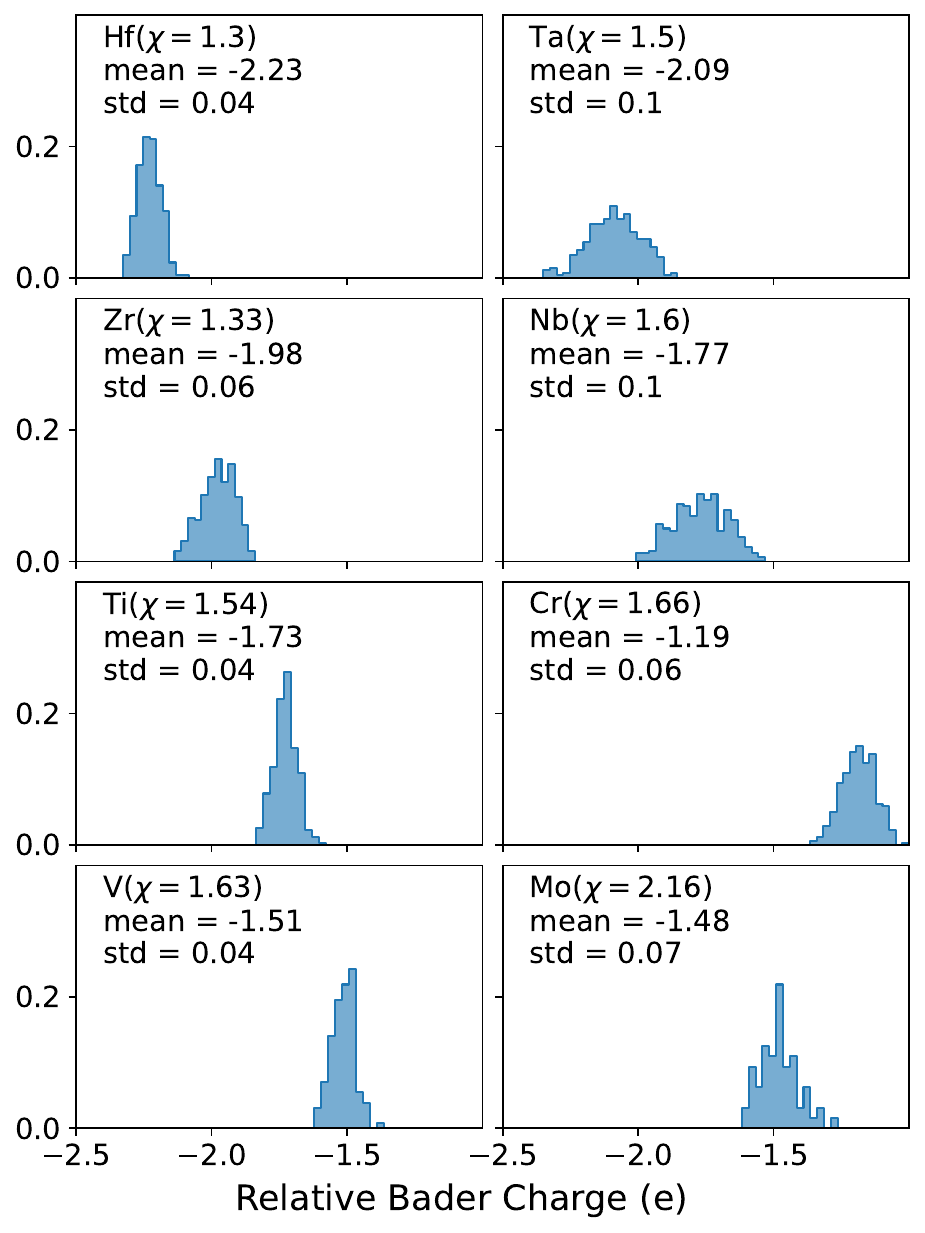}
\caption{
Probability histograms of the relative Bader charge for the eight metals in all of our pristine HEMN supercells.
The Pauling electronegativity ($\chi$), mean relative Bader charge, and standard deviation of the relative Bader charge are included. 
The Bader charge is relative to the number of valence electrons of the metal.
}
\label{fig:BaderCharge}
\end{figure}

The trend we observe in our study of nitrogen environments with high or low vacancy formation energy, characterized by similar metallic compositions, correlates with findings from other studies on high-entropy carbides, some of which suggest that electronegativity is a key property that can help predict vacancy formation energy.
Zhao \textit{et al.}\ found that the effect of increasing carbon vacancy formation energy in (ZrHfNbTa)C increases with adding more Nb, Ta, Zr, and Hf to the local environment in increasing order \cite{zhao2023machine}.  
Similarly, Lu \textit{et al.}\ found that carbon vacancy formation energy in (TiZrHfNb)C increases with adding more Nb, Ti, Zr, and Hf in increasing order \cite{lu2025coupling}.
This result is interesting since the coefficients in our nn-model in Table~\ref{tab:Model_coefs} show that nitrogen vacancy formation energy favors those same metals in the same order. 
Lu \textit{et al.} \ also states that metals with lower electronegativity have greater charge transfer with carbon, resulting in enhanced bond stability which makes creating vacancies unfavorable \cite{lu2025coupling}.
This trend can be seen with the metals they studied---Nb, Ti, Zr, and Hf---whose electronegativities are 1.6, 1.54, 1.33, and 1.3.
While this trend is clear in Lu \textit{et al.}'s work, the electronegativity for the metals in our HEMNs does not fully predict the charge transfer; furthermore, neither the charge transfer nor electronegativity fully correlates with the vacancy formation energy.
Fig.~\ref{fig:BaderCharge} shows the distribution of charge transfer (through relative Bader charge) for each of the metals in our HEMNs where Hf and Zr lose more electrons to their neighboring nitrogen atoms than Mo and Cr. This also shows that the amount of charge transfer from the metal to the nitrogen is not strictly predicted by the electronegativity; moreover, neither the electronegativity alone nor the charge transfer is sufficient to predict the effect on nitrogen vacancy formation energies.
Despite this, we see other similarities between high entropy carbides and nitrides; for example, the ordering of these metals in their respective high-entropy carbides aligns with the vacancy formation energy trends found in the corresponding binary carbides \cite{lu2025coupling}, which is also the general trend we found with HEMN and binary nitride vacancy formation energy.
This shows that there may be some similarities between how carbon and nitrogen interact with their nearest-neighbor environment. 
This suggests that the ordering of nitrogen vacancy formation may also hold for high-entropy carbides and other high-entropy ceramic materials. 

It is clear that the composition of the anion local environment in high-entropy carbides, oxides, and nitrides is vital in predicting the vacancy formation energy; moreover, we find that including features beyond the 1nn shell provides little benefit to model effectiveness. 
Both Zhao \textit{et al.}\ and Lu \textit{et al.}\ included features up to the fifth nearest-neighbor shell of metals around a carbon vacancy in their model feature sets \cite{zhao2023machine,lu2025coupling}.
Both studies cite that including up to the fifth nearest neighbors provides a better MAE in their model; however, this improvement is generally minimal.
Zhao \textit{et al.}\ found improvement in MAE for the testing set between 1nn and 5nn shell features is only 0.018 eV \cite{zhao2023machine}.
The paper by Lu \textit{et al.}\ finds an improvement in MAE is on the order of around 0.06 eV by including 5nn shell features compared to 1nn features \cite{lu2025coupling}.
While the MAE improves when using features up to the 5nn shell compared to the 1nn shell, these studies do not discuss how much of this improvement is due to increased model complexity compared to goodness of fit, as showing differences in MAE do not effectively describe this.
Additionally, with improvements in the MAE on the order of 0.06 eV and lower, we can see that going beyond the 1nn shell metals can only provide a small correction to the value predicted from just the 1nn environment.
We find that if we create a model that includes both 1nn and 3nn features, the improvement we get in the RMSE for HEMNs is only 3 meV; furthermore, the Bayesian Information Criteria (BIC) for our original nn-model is 0.19 while the BIC for the model that includes 3nn features is 28.88, showing that our nn-model is the more optimized model. 
Furthermore, the BIC for our triplet model is 15.62, showing that the best way to improve our original nn-model is to add more degrees of freedom within the first nearest-neighbor shell than using features in the third nearest-neighbor shell.
This effect can be further explained through Fig.~\ref{fig:Delta_EDM_BC}, where we see that the EDM energy of atoms in the 2nn and 3nn shell around a vacancy are not changed nearly as much as the 1nn shell atoms, providing a physical explanation for the low change in error when using features beyond the 1nn in the models found by Zhao \textit{et al.}\ , Lu \textit{et al.}\ , and our own.

The other studies of vacancy formation energies in high-entropy ceramics all use the special quasi-random structure (SQS) method to create their vacancy environments \cite{zhao2023machine,lu2025coupling,chae2022effects}; on the other hand, our study uses both a different algorithm to create our supercells as well as using EDM to screen high and low EDM nitrogen sites to create vacancies.
Zhao \textit{et al.}\ created 40 SQS supercells and calculated the vacancy formation energy on every carbon site resulting in 1280 carbon vacancies \cite{zhao2023machine}, Lu \textit{et al.}\ created 6 SQS for (TiZrHfNb)C and created 14 anion vacancies per SQS to calculate vacancy formation energy \cite{lu2025coupling}, while Chae \textit{et al.}\ created 1 SQS for (MgCoNiCuZn)O and calculated the vacancy formation energy for 40 oxygen vacancies in that supercell \cite{chae2022effects}.
While the SQS method may provide a variety of random anion environments, the method does not necessarily maximize the number of unique environments found throughout the set of supercells created.
The paper by Zhao \textit{et al.}\ details that the compositions of their carbon environments in their 40 SQS supercells match closely with the true random high-entropy carbide \cite{zhao2023machine}; however, they do not detail how many of those compositions are symmetrically unique, while also showing that even with 40 supercells they still have missing environments. 
Our algorithm ensures that any set of supercells created with it will be optimized concerning the number of unique anion environments, allowing for a more optimized set of supercells for anion vacancy formation energy purposes in B1 high-entropy ceramics over the SQS method.
Additionally, using EDM energies to select nitrogen sites for vacancies provides a systematic approach to find environments with the lowest (or highest) vacancy formation energies. 
Using our algorithm to create supercells while using EDM to select nitrogen sites for vacancies gives a more efficient database of formation energies while remaining  representative. 

\begin{figure}[htbp!]
\includegraphics[width=3.3in]{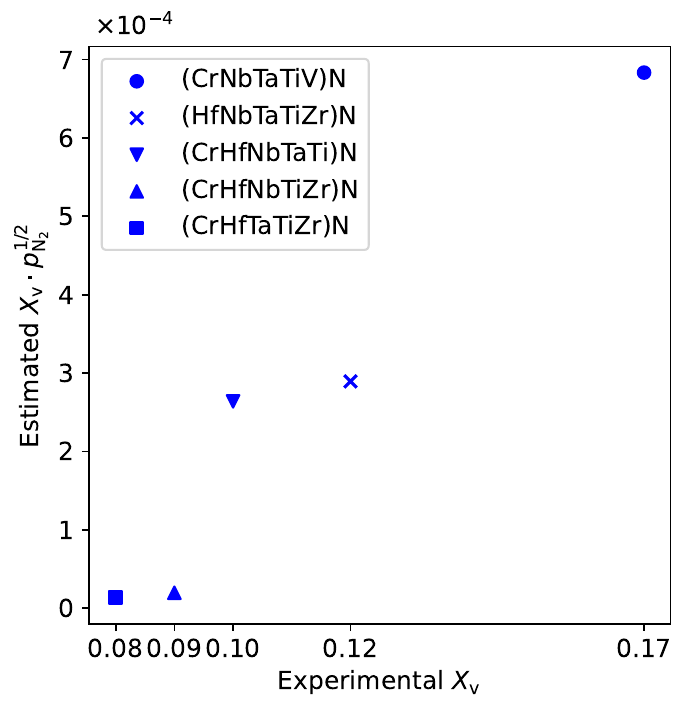}
\caption{
Reaction rate equilibrium estimate of the nitrogen vacancy concentration with the nitrogen partial pressure during synthesis ($p_{\text{N}_{2}}$) calculated using coefficients from our nn-model against experimental HEMN vacancy concentrations taken from the HEMN stoichiometries from Dippo \textit{et al.}\ \cite{dippo2020bulk}.
The nitrogen partial pressure affects the nitrogen vacancy concentration; it is unknown, but expected to be small when sintering in an argon atmosphere.
}
\label{fig:Nvac_concentration}
\end{figure}

Fig.~\ref{fig:Nvac_concentration} shows a comparison with our model vacancy energies and the experimental nitrogen concentrations in HEMNs.
Dippo \textit{et al.}\ synthesized bulk (HfNbTaTiZr)N, (CrNbTaTiV)N, (CrHfNbTaTi)N, (CrHfNbTiZr)N, and (CrHfTaTiZr)N at near stoichiometric ratios of metal to nitrogen in the B1 crystal structure, followed by high-temperature sintering \cite{dippo2020bulk}.
The reaction equation for nitrogen vacancy formation, $\text{N}_{\text{N}} \rightleftharpoons \text{V}_{\text{N}} + \frac{1}{2}\text{N}_{\text{2}}(\text{g})$, and the resulting equilibrium constant given by the law of mass action formula, $k = X_{v}p_{\text{N}_2}^{1/2} / \left[ \text{N}_{\text{N}} \right]$, show that the vacancy concentration $X_{v}$ varies with nitrogen gas partial pressure $p_{\text{N}_{\text{2}}}$.
While the exact value of the partial pressure is unknown, the synthesis of the HEMNs was done in an argon glove box \cite{dippo2020bulk}, so we can assume that the partial pressure of nitrogen is low.
Additionally, Dippo \textit{et al.}\ state that each HEMN has about 4--10 at.\% carbon, which should occupy nitrogen sites and displacing nitrogen from nitrogen sites. 
The equilibrium constant $k$ is averaged over nitrogen environments, with the 5 possible metals in each of the 6 nearest-neighbor sites represented as $\alpha_{1}$ to $\alpha_{6}$,
\begin{equation*}
    X_v \cdot p^{1/2}_{\text{N}_\text{2}}  = f_{\text{N}}\sum_{\alpha_1, \ldots, \alpha_6}  \prod_{i=1}^{6} f_{\alpha_i} \exp\left(-(k_\text{B}T_\text{sintering})^{-1} \sum_{i=1}^{6} E_{vf,\alpha_i}\right).
\end{equation*}
The fraction of each metal in the HEMN, $f_{\alpha_{i}}$, comes from the EDS values from Dippo at al. \cite{dippo2020bulk}, along with the fraction of nitrogen in anion sublattice $f_{\text{N}}$, and the sintering temperature $T_\text{sintering}$. 
The nn-model coefficients for the constituent metals are $E_\text{vf,$\alpha_{i}$}$. 
We plot $X_{v}p_{\text{N}_2}^{1/2}$ against the experimental vacancy concentration from Dippo \textit{et al.}\ in Fig.~\ref{fig:Nvac_concentration} and find that our estimated vacancy concentration correlates with experimental results and provides a qualitative prediction of true vacancy concentrations in bulk HEMNs.
While we don't know $p_{\text{N}_2}$, we assume that it is low, and similar for the 5 nitrides.
However, the experimental vacancy concentrations range from 0.08--0.17, which would exceed the dilute limit approximation we used with the law of mass action.
To further improve our estimate, experimental activity coefficients for vacancy interaction would be needed.
All our DFT calculations were done with the PBE functional; however, we do not expect using other functionals like a meta-GGA would change the accuracy of our estimated vacancy concentration in Fig.~\ref{fig:Nvac_concentration}.
The study by Friedrich \textit{et al.} compared the formation enthalpies of rocksalt TiN, VN, CrN, ZrN, NbN, HfN, and TaN calculated with both PBE and the SCAN meta-GGA functionals, and found that they have nearly the same average error \cite{friedrich2024aflow}.
As such, we do not expect the difference in the vacancy formation energy to be significant with a meta-GGA as compared to PBE.
Additionally, while using a different functional may change the actual value of the calculated vacancy formation energy, we do not expect the ordering of the resulting coefficients from our model to change drastically.
Therefore, while using a meta-GGA or not should not change the qualitative prediction of whether or not a bulk HEMN will have a higher or lower vacancy concentration.
The results in Fig.~\ref{fig:Nvac_concentration} validate our model against experimental results, but also show that there is additional information that is needed to quantitatively predict vacancy concentrations that is not covered in our estimate.

\section{Conclusions}

In this work, we created a set of HEMN supercells that maximize the number of unique nitrogen environments we can simulate in order to quantify the effects of local environment on the energetics of nitrogen vacancies in HEMNs.
We built a set of 10 supercells in which 97\% of the nitrogen environments are symmetrically unique and we relaxed each set of supercells in order to keep the cell size the same for each HEMN.
The EDM energies of nitrogen in binary, ternary, and HEMN systems correlate similarly with each other, detailing the effects of local environment on the energy density of a nitrogen atom.
Using our EDM data we qualitatively predicted which HEMN nitrogen sites would have high and low vacancy formation energy.
We built a simple weighted linear regression model to effectively predict nitrogen vacancy formation energy in HEMNs, and extended this model to extend to ternary and binary nitride systems.
We contrasted with a model built on binary nitride vacancy formation energy data, which had poor estimation of ternary and HEMN vacancy formation energies, showing the effect of training environment. 
Finally, the EDM for vacancy environments showed the largest changes in the first nearest-neighbor shell. This both explains why EDM alone had difficulty in predicting the vacancy formation energy, and why a model built from the first neighbor composition could be accurate.
Triplet features to our nn-model gave a slight improvement to our predictions as well as providing additional information on how our nn-model predicts the vacancy formation energy of various nitrogen environments.
While EDM was not as strong a predictor for vacancy formation energy, it may still be useful when combined with other features that quantify the local environment, or in differentiating local environments with the same nearest-neighbor composition. 
It may be possible to build future models that predict the change in EDM energies when forming a defect, that may allow better predictions of the vacancy formation energy directly from EDM calculations alone for HEMNs and other high-entropy ceramics. 
Furthermore, the data from our calculations can be used to construct other models, as well as guide experimental studies on how the composition of various HEMNs can affect the nitrogen vacancy concentration in those systems, to build more effective HEMNs coatings for high hardness applications. 

\section{Acknowledgments}

The authors acknowledge research support from the Laboratory Directed Research and Development (LDRD) program at Sandia National Laboratories. Sandia National Laboratories is a multimission laboratory managed and operated by National Technology \& Engineering Solutions of Sandia, LLC, a wholly owned subsidiary of Honeywell International Inc., for the U.S.~Department of Energy's National Nuclear Security Administration (DOE/NNSA) under contract DE-NA0003525. This written work is authored by an employee of NTESS. The employee, not NTESS, owns the right, title and interest in and to the written work and is responsible for its contents. Any subjective views or opinions that might be expressed in the written work do not necessarily represent the views of the U.S. Government. The publisher acknowledges that the U.S. Government retains a non-exclusive, paid-up, irrevocable, world-wide license to publish or reproduce the published form of this written work or allow others to do so, for U.S. Government purposes. The DOE will provide public access to results of federally sponsored research in accordance with the DOE Public Access Plan. The data is available at the Materials Data Facility \cite{Blaszik2016, Blaszik2019}, doi:10.18126/sadd-1e78 \cite{DeSilva2025data}.


%
\end{document}